\documentclass{article}
\usepackage[square,numbers]{natbib}
\usepackage{authblk}
\usepackage{charter}
\usepackage{fullpage}
\usepackage{bbm}

\usepackage{hyperref}       
\usepackage{url}            
\usepackage{booktabs}       
\usepackage{amsfonts}       
\usepackage{nicefrac}       
\usepackage{microtype}      
\usepackage{xcolor}         
\usepackage{amsmath,amssymb,amsthm,amsfonts}
\usepackage{latexsym,bbm,graphicx,float,mathtools}
\usepackage[export]{adjustbox}
\usepackage{capt-of}
\usepackage{booktabs}
\usepackage{parskip}

\usepackage[utf8]{inputenc}

\usepackage[noabbrev]{cleveref}
\usepackage{caption}
\usepackage{subcaption}
\usepackage{transparent}
\usepackage[inline]{enumitem}
\usepackage{multirow}
\usepackage{multicol}
\usepackage{algorithm}
\usepackage{algpseudocode}
\usepackage{xspace}

\usepackage{abstract}

\usepackage[subtle, mathdisplays=tight, charwidths=normal, leading=normal]{savetrees}

\title{\textbf{Zero-Indexing Internet Search Augmented Generation for Large Language Models}}

\author{
    Guangxin He$^{*\dag{\S}}$, Zonghong Dai$^{*\ddag}$, Jiangcheng Zhu$^{\S}$, Binqiang Zhao$^{\S}$, Qicheng Hu$^{\S}$\\
    \vspace{-0.75em}
    Chenyue Li$^\dag$, You Peng$^\dag$, Chen Wang$^\Vert$, Binhang Yuan$^\dag$\\
    \dag \ HKUST \ \ \ \ \ddag \ Fudan University \ \ \ \ ${\S}$ 01.AI \ \ \ \ $\Vert$ Tsinghua University
}

\date{}

\begin{document}
\maketitle

\vspace{-2ex}
\begin{abstract}
Retrieval augmented generation has emerged as an effective method to enhance large language model performance. This approach typically relies on an internal retrieval module that uses various indexing mechanisms to manage a static pre-processed corpus. However, such a paradigm often falls short when it is necessary to integrate the most up-to-date information that has not been updated into the corpus during generative inference time. In this paper, we explore an alternative approach that leverages standard search engine APIs to dynamically integrate the latest online information (without maintaining any index for any fixed corpus), thereby improving the quality of generated content. 
We design a collaborative LLM-based paradigm, where we include: (\underline{i}) a \textsc{parser-LLM} that determines if the Internet augmented generation is demanded and extracts the search keywords if so with a single inference; (\underline{ii}) a \textit{mixed ranking strategy} that re-ranks the retrieved HTML files to eliminate bias introduced from the search engine API; and (\underline{iii}) an \textsc{extractor-LLM} that can accurately and efficiently extract relevant information from the fresh content in each HTML file. 
We conduct extensive empirical studies to evaluate the performance of this Internet search augmented generation paradigm. The experimental results demonstrate that our method generates content with significantly improved quality. Our system has been successfully deployed in a production environment to serve 01.AI's generative inference requests.
\end{abstract}



\section{Introduction}
\label{sec:intro}
\vspace{-0.5em}

Large language models (LLM) have demonstrated remarkable capabilities, fueling a new wave of innovative AI applications~\cite{bommasani2021opportunities}. To incorporate information not included during their training phase, retrieval augmented generation (RAG) has been developed as an efficient method to enhance LLM performance by integrating externally retrieved information~\cite{zhao2024retrieval,ding2024survey}. Usually, RAG systems use an internal retrieval module that employs various indexing mechanisms to manage a static corpus. In this paper, we study an alternative approach, exploring the design and implementation of \textit{a RAG paradigm that leverages standard search engine APIs} (such as Google and Bing), which allows for the flexible and dynamic integration of the most update-to-date online information, thereby improving the quality of content generated by LLMs.

While RAG systems with a static retrieval component, which leverages a fixed corpus of data, are effective for tasks within well-defined knowledge domains, Internet search augmented generation~\cite{nakano2021webgpt,komeili2022internet,lazaridou2022internet,li2023web,duffy2023dotori,xie2024weknow} offers distinct advantages. By accessing the \textit{most update-to-date information} available online, internet search augmented generation is particularly beneficial for tasks that require up-to-date data, such as news updates, market trends, or recent scientific discoveries, enabling the model to produce more relevant and timely responses~\cite{xie2024weknow}. Moreover, Internet search augmented systems can retrieve information from a vast array of sources across the web, encompassing a \textit{broader range} of topics and perspectives, thereby enhancing the comprehensiveness and diversity of the generated content~\cite{lazaridou2022internet}. In fact, OpenAI has also released its toolkit \textsc{ChatGPT-Search} to integrate information from the Internet to improve the user experience very recently~\cite{chatgpt_search}.

Building a production-level, general-purpose Internet search augmented generation system is a complex and challenging task. Unlike standard RAG systems that rely on static retrieval methods—such as sparse~\cite{salton1988term,robertson2009probabilistic,paulsen2023sparkly}, dense~\cite{wang2024multilingual,chen2024m3,lee2024nv}, or more advanced learnable indexing~\cite{wang2022lider,zhang2022plin,zhao2023towards,sun2023learned,li2023dili,zeakis2023pre,lampropoulos2023adaptive,zhang2024hyper} mechanisms—to retrieve carefully pre-processed, chunked texts stored as key-value pair from a vector database~\cite{cai2023bonsaikv,zhang2023experimental,lu2024fluidkv,yu2022treeline,ahmad2022pantheon,wang2023mirrorkv,mo2023learning,su2024vexless} for context generation, Internet search augmented generation must \textit{dynamically} process content returned by search engine APIs --- otherwise any platform that serves the generative inference requests has to re-build the search engine infrastructure of Google or Bing from scratch internally. The Internet search augmented generation paradigm requires an approach that is not only economically viable but also highly efficient, capable of quickly extracting relevant information from large-scale unstructured data from the web to feed the LLM prompt input. The real-time nature of this processing, combined with the need to accurately extract relevant information across diverse and constantly changing content, significantly increases the complexity of deploying such a system at the production scale.

Naive approaches often fall short when attempting to build a production-level Internet search augmented generation system. The raw data retrieved from search engine APIs is usually highly redundant and noisy, which can degrade the quality of the content generated by LLMs. For instance, a simplistic approach might involve applying a set of rule-based text extraction heuristics; however, this method is labor-intensive and struggles to generalize across the diverse and ever-changing landscape of online information.
Moreover, despite the advancements in LLMs that now support significantly longer context lengths --- such as \textsc{GPT-4o}~\cite{gpt4o} and \textsc{Llama 3.1}~\cite{dubey2024llama}, which handle up to 128K tokens, and \textsc{Yi}~\cite{young2024yi} and \textsc{Claudi3}~\cite{claude3}, which can process 200K tokens --- feeding raw, unprocessed data directly from search engines into these models is inefficient --- such approach not only increases the computational load, but also risks diminishing the quality of the generated content~\cite{hsieh2024ruler}. To truly harness the potential of Internet search augmented generation, more sophisticated techniques are needed to filter and process online information before it is used in LLM content generation.

Towards this end, we design and implement an Internet search augmented generation system deployed in 01AI's production-level inference API system. Concretely, we summarize our key contributions:

\vspace{0.15em}
\textbf{\underline{Contribution 1.}}  We propose a novel zero-indexing paradigm for Internet search augmented generation, where we introduce a set of core components to collaboratively accomplish the task. Concretely, we implement (\underline{i}) a \textsc{parser-LLM} that determines the necessity of Internet-augmented generation and, if required, extracts search keywords in a single inference pass; (\underline{ii}) a \textit{mixed ranking strategy} that that re-ranks the retrieved HTML files to mitigate biases introduced by the search engine API; and (\underline{iii}) an \textsc{extractor-LLM} that is designed to efficiently and accurately extract relevant information from the fresh content within each HTML file. Unlike traditional methods that rely on indexing or storage modules, our approach retrieves all augmented information via the Google and Bing Search APIs and processes it on the fly, eliminating the need to manage the pre-indexed data. This paradigm allows for real-time, contextually relevant content generation without the overhead of maintaining a static corpus, making it adaptable for a wide range of applications.

\vspace{0.15em}
\textbf{\underline{Contribution 2.}} We enumerate the design principle and implementation details for the core components: (\underline{i}) we introduce how we construct the instruction set to train the \textsc{parser-LLM} to equip it with multi-instruction following ability and be able to extract the keywords for the search engine API precisely; (\underline{ii}) we implement the \textit{mixed ranking strategy} that considers both the snippet and the full content in HTML file to provide more effective re-ranking based on pure semantic relevance; (\underline{iii}) we apply both supervised fine-tuning and direct preference optimization to train the \textsc{extractor-LLM} so that it can accurately extract relevant information from each retrieved HTML file to improve the end-to-end generation quality. 

\vspace{0.15em}
\textbf{\underline{Contribution 3.}} We conduct a set of concrete empirical evaluations to estimate the design and implementation of our search augmented generation system. The experimental results suggest that: (\underline{i}) our Internet search augmented generation paradigm generates higher-quality outputs when compared with other RAG paradigms, which is also more cost-efficient in terms of the processed input tokens for the \textsc{generative-LLM}; (\underline{ii}) the \textsc{extractor-LLM} can extract the relevant information accurately and precisely for various benchmarks, where it can also robustly reject inappropriate content rather than generate hallucinations when there is no relevant content. 
    

\begin{figure*}[t]
    \vspace{-1ex}
    \subfloat[Standard retrieval augmented generation.]{
        \label{fig_max_model_size_motivation}
        \includegraphics[width=0.5\linewidth]{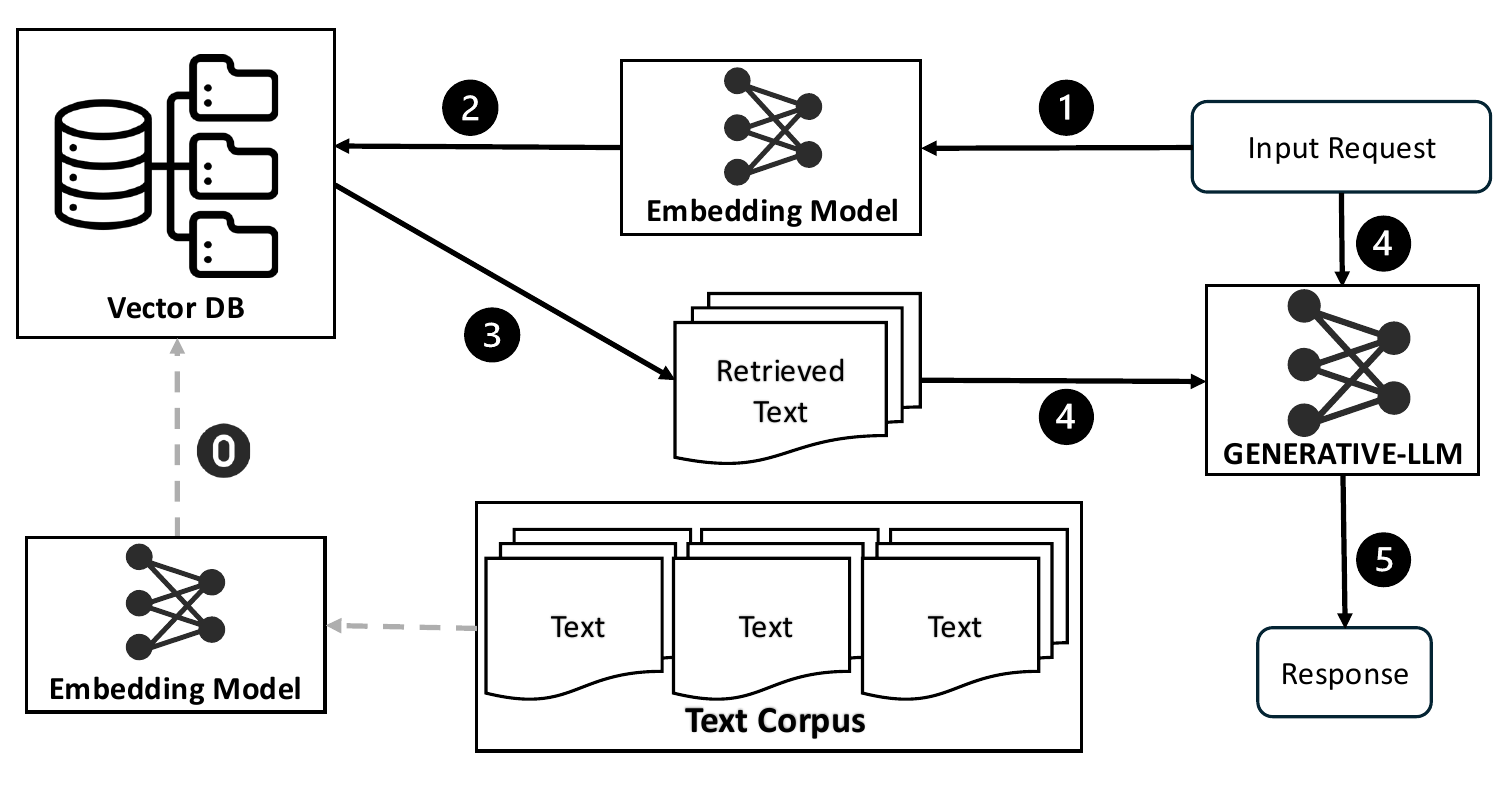}
    }
    \hfill    
    \subfloat[Internet search augmented generation.]{
        \label{fig:overall_gpu_util}
        \includegraphics[width=0.5\linewidth]{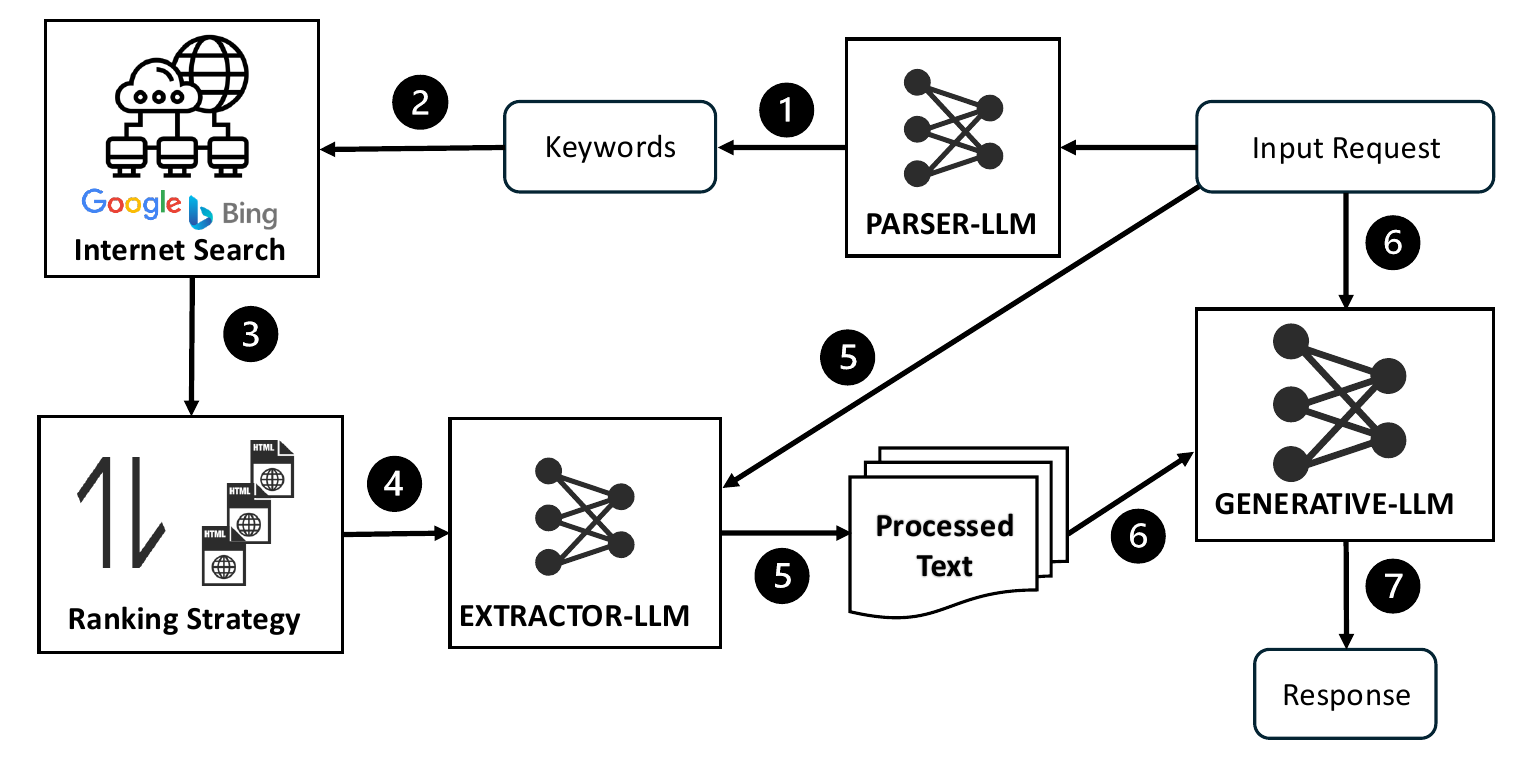}
    }
\vspace{-1ex}
\caption{ A comparison between the standard RAG paradigm and our Internet search augmented generation paradigm.} 
    \label{fig:framework} 
\vspace{-4ex}
\end{figure*} 



\section{Internet Search Augmented Generation}
\label{sec:isag}
\vspace{-0.5em}

We first introduce our proposed framework to support the paradigm of Internet search augmented generation. Figure \ref{fig:framework} compares the standard RAG and the Internet search augmented generation proposed in this paper. 

\subsection{Paradigm Overview}
\label{sec:parad_overview}

Before we introduce our Internet search augmented generation, we first briefly review the paradigm of standard RAG as we illustrated in Figure \ref{fig:framework}(a). Before processing any inference request, standard RAG needs to prepare key-value pairs that need to be stored in a vector database, where a value is a piece of chunked text from the corpus, and the key is the corresponding embedding generated by the embedding model (\textbf{Step 0} in Figure \ref{fig:framework}(a)). When an inference request arrives, the following steps will be conducted:

\vspace{-0.5em}
\begin{itemize}[topsep=5pt, leftmargin=1em]
    \item \textbf{Step 1}. The RAG system utilizes an embedding model (typically the same model used during the indexing phase) to generate an embedding vector for the input request.
    \vspace{0.15em}
    \item \textbf{Step 2}. This embedding vector is then used to query the vector database created during the indexing phase to search for semantically similar text chunks.
    \vspace{0.15em}
    \item \textbf{Step 3}. The RAG system retrieves and selects the most relevant text from the vector database based on some similarity measurements (e.g., cosine similarity).
    \vspace{0.15em}
    \item \textbf{Step 4}. The RAG system combines the original input request with the retrieved text to form a new prompt that integrates additional information.
    \vspace{0.15em}
    \item \textbf{Step 5}. The generative LLM uses this new prompt to generate a response, leveraging the retrieved information to produce more accurate and contextually relevant output.
\end{itemize}
\vspace{-0.5em}

By contrast, our Internet search augmented generation works in a fully dynamic paradigm by interacting with the external search engine APIs, which requires \textit{zero-indexing} mechanism. Figure \ref{fig:framework}(b) illustrates our proposed paradigm when processing an inference request:  

\vspace{-0.5em}
\begin{itemize}[topsep=5pt, leftmargin=1em]
\item \textbf{Step 1}. Our system first leverages a multi-functional \textsc{parser-LLM} to determine whether the input request requires external information from the Internet to generate the response --- If external information is needed, the \textsc{parser-LLM} will interpret and extract the keywords based on the input request, which will be provided to the search engine API. Note that these functionalities are completed simultaneously within a single inference pass by the \textsc{parser-LLM}.

\vspace{0.15em}
\item \textbf{Step 2}. The parsed keywords are then used to query external search APIs, such as Google or Bing, to retrieve relevant information from the Internet.
\item \textbf{Step 3}. We employ a mixed ranking strategy that incorporates two levels of granularity to rerank retrieved HTML files so that the original bias introduced by the search engines can be reduced.

\vspace{0.15em}
\item \textbf{Step 4}. We apply some lightweight pre-processing to insert content tags for the retrieved text so that it can be further processed by the \textsc{extractor-LLM}. 

\vspace{0.15em}
\item \textbf{Step 5}. As another core component of our system, the \textsc{extractor-LLM} first analyzes the original input request along with the tagged retrieved content, then extracts the most relevant text portion identified by tags for generation. 

\vspace{0.15em}
\item \textbf{Step 6}. The system combines the original input request with the extracted text to form an enriched prompt that integrates the additional information from the Internet.

\vspace{0.15em}
\item \textbf{Step 7}. The backbone \textsc{generative-LLM} uses this new prompt to generate a response, leveraging the information from the Internet to produce high-quality output with the potential most up-to-date information.  
\end{itemize}
\vspace{-0.5em}

When compared with the standard RAG system, the Internet search augmented generation needs to answer the following two essential questions:

\begin{quote}
    \textit{\textbf{Question 1}. How can we accurately interpret the user's intent to interact with the search engine API, where we dynamically search the Internet and re-organize the retrieved context?}
\end{quote}

\begin{quote}
    \textit{\textbf{Question 2}. How can we accurately and efficiently extract contextually relevant information to improve the quality of the generated output by implementing the \textsc{extractor-LLM}?} 
\end{quote}

\noindent To answer these questions, we first identify the key challenges in Section \ref{sec:challenges}, then we propose our system design and implementation to overcome these challenges in Section \ref{sec:design}.

\subsection{Key Challenges}
\label{sec:challenges}

Given the Internet search augmented generation paradigm, we identify two critical challenges, each addressed by dedicated modules outlined in Section \ref{sec:design}:

\underline{\textbf{Challenge 1.}} \textit{Effectively retrieving intent-aligned web pages from the Internet without biased ranking.}
Effectively retrieving web pages that align with user intent involves solving two critical sub-problems: (\underline{i}) how to interpret user requests comprehensively before the search; and (\underline{ii}) how to mitigate biases in search result rankings after the search. To be more specific, we summarize the key issues below: 

\vspace{-0.5em}
\begin{itemize}[topsep=5pt, leftmargin=1em]
    \item \textbf{Comprehensive interpretation of user requests}: a critical module in our system (i.e., the \textsc{parser-LLM}) must accurately understand and interpret user queries in a way that captures the full scope of the user's intent. Ambiguities in user requests, variations in language, and the context-dependent nature of search queries create substantial obstacles. Although some recent RAG systems have implemented a similar component in question answering~\cite{lazaridou2022internet,levonian2023retrieval} and query rewriting mechanisms~\cite{ma2023query,su2024dragin,chan2024rq}, accurately inferring the user's true intent without oversimplification or misinterpretation remains a persistent challenge in our paradigm.

    \vspace{0.15em}
    \item \textbf{Mitigating bias in search result rankings}: ranking bias in search engines remains an obstacle in our paradigm since reliance on engagement metrics such as click-through rate (CTR)~\cite{richardson2007predicting} and dwell time~\cite{yi2014beyond} often skews result rankings, which prioritize user interaction over the actual relevance of the content --- such a ranking bias needs to be removed to improve the generation quality.
\end{itemize}

\vspace{-0.5em}
\underline{\textbf{Challenge 2.}} \textit{Accurately and efficiently extracting contextually relevant contents from each retrieved HTML file.} When implementing the \textsc{extractor-LLM}, we need to fulfill the following two concrete demands:

\vspace{-0.5em}
\begin{itemize}[topsep=5pt, leftmargin=1em]

\item \textbf{Accuracy in content extraction}: to ensure that the information retrieved from the internet is genuinely beneficial for output generation, it is crucial that the extracted content is \textit{contextually relevant to} the original input request --- in our paradigm, this indicates that the \textsc{extractor-LLM} must accurately identify and select the most pertinent parts of the retrieved data, avoiding the inclusion of irrelevant or misleading content, which is particularly challenging given the variability and complexity of web data.

\vspace{0.15em}
\item \textbf{Computational efficiency in context extraction}: since large volumes of HTML data must be processed rapidly to maintain real-time response generation when serving the inference API, we need to guarantee computational efficiency of the \textsc{extractor-LLM} in our paradigm. Therefore, it is essential to have a lightweight LLM for this functionality, which ensures the system can deliver fast, accurate, and scalable performance by leveraging economic computational resources.
\end{itemize}

\subsection{Core Functionality Design Overview}
\label{sec:design}
To implement the paradigm introduced in Section~\ref{sec:parad_overview} and address the two challenges outlined in Section~\ref{sec:challenges}, we provide an overview of the design of the two associated functionalities:

\vspace{0.15em}
\textbf{Intent-alignmented Internet Search.}
To obtain well-processed Internet search results, i.e., reranked HTML files, for further processing, we need to integrate the \textsc{parser-LLM}'s functionality from \textbf{Step 1} in Figure~\ref{fig:framework}(b) and incorporate the reranking mechanism described in \textbf{Step 3} in Figure~\ref{fig:framework}(b). To address the challenges of effectively retrieving intent-aligned web pages, we propose a hybrid approach that combines keyword extraction with reranking strategies. Concretely, the process begins with the \textsc{parser-LLM}, which is carefully trained to achieve two primary objectives when preparing user requests for search: (\underline{i}) to comprehensively understand the user's intentions and (\underline{ii}) to extract and output these intentions in the form of a list of keywords, allowing for optimal utilization of search engines. 
After retrieving search results using the Parser LLM, we implement a \textit{mixed ranking strategy} that combines fine-grained with coarse-grained content ranking. This approach mitigates biases influenced by commercially driven engagement factors. The strategy combines with our prior keyword extraction methodology, ensuring the effectiveness of the internet search retrieval. We enumerate the design and implementation details of this functionality in Section~\ref{sec:internet_search}.

\vspace{0.15em}
\textbf{Content Extraction from Search Results.} Given the well-organized Internet search results, we need to efficiently and accurately extract relevant information from the retrieved HTML files; our approach involves a two-step processing pipeline: (\underline{i}) pre-processing the HTML content, and (\underline{ii}) identifying relevant information using the \textsc{extractor-LLM} with efficient computation cost during extraction. The preprocessing of the HTML files leverages a simple finite state machine to segment the content in the HTML file, enclosing each segment within corresponding tags. 
To guarantee the extraction accuracy, we implement a two-phase strategy to train the \textsc{extractor-LLM}: (\underline{i}) the first phase uses supervised fine-tuning to train the model's basic ability to extract relevant information with careful construction of the instruction set, and (\underline{ii}) the second phase further aligns the extraction with the preference of a judge after training with reinforcement learning technique. 
Additionally, to enhance computational efficiency, we optimize the \textsc{extractor-LLM} in two ways: (\underline{i}) we choose a relatively light-weight 9B LLM, where the inference latency is low; (\underline{ii}) we only require the \textsc{extractor-LLM} to generate relevant tags instead of directly output the original segmented contents, which significantly decreases the number of output tokens for the \textsc{extractor-LLM}, and thus accelerate the extracting processing time. This integrated process ensures that the information extraction is both accurate and efficient, enabling faster and more effective retrieval of relevant information from segmented HTML content. The detailed discussion about how we implement this functionality is illustrated in Section~\ref{sec:train}. 



\section{Retrieval from Search Engines}
\label{sec:internet_search}

As we introduced in Section \ref{sec:isag}, effectively organizing the Internet search results is indispensable in our Internet search augmented generation paradigm. In pursuit of the goal of user intent alignment and elimination of the bias introduced by the search engine APIs, we apply the following design: we introduce our training procedure of the \textsc{parser-LLM} in Section \ref{sec:parser} to refine the user's query before Internet search and the \textit{mixed ranking strategy} in Section \ref{sec:ranker} to optimize the ranking of the results returned from search engine APIs. 

\subsection{Training the \textsc{parser-LLM}}
\label{sec:parser}

Functionally, the \textsc{parser-LLM} takes the original user \texttt{request} as input, where two primary tasks need to be accomplished by a single inference computation: (\underline{i}) the \textsc{parser-LLM} assesses whether Internet search is necessary to improve the generation quality based on the user query prompt;  (\underline{ii}) If deemed necessary, \textsc{parser-LLM} need to extract a list of effective keywords from the user query prompt, where during the keyword extraction workflow, the \textsc{parser-LLM} needs to solve the potential issues of time-sensitivity, single-language limitation, and out-of-vocabulary error. Concretely, we enumerate the following problem that needs to be resolved by the training strategy design:  

\vspace{-0.5em}
\begin{itemize}[topsep=5pt, leftmargin=1em]

    \item \textbf{Multi-instruction following}: in order to improve the system's efficiency, we demand the \textsc{parser-LLM} to generate the indication of whether Internet search is demand and the corresponding extracted keywords in a single inference computation, the \textsc{parser-LLM} has to be able to interpret the inherited relationship between the two tasks and generate integrated results.    
    
    \vspace{0.15em}
    \item \textbf{Time information sensitivity}: the effectiveness of the keyword extraction can vary significantly depending on how time-specific information is integrated into the extraction prompt, where the impact varies based on different type of \texttt{request}. For instance, in cases with positive temporal dependency (e.g., "How's the weather?"), adding explicit time details to the keywords improves search accuracy by narrowing the context. However, for other queries (e.g., "What is the best weight loss method of 2024?"), including temporal information such as "2024" might inadvertently reduce search performance by limiting the scope of relevant results during the search process.

    \vspace{0.15em}
    \item \textbf{Monolingual search limitation}: relying solely on monolingual keyword extraction, such as the language of the user prompts, may not fully capture the essential keywords, especially when dealing with high-context and deductive \texttt{request}. For example, some languages are concise and rich in meaning but prone to ambiguity, making it challenging for the \textsc{parser-LLM} to extract the keywords~\cite{Chen_Xu_Li_Yu_Pan_Zhang_2024}. This limitation highlights the difficulty of interpreting complex user queries and retrieving relevant results effectively when confined to a single language.

    \vspace{0.15em}
    \item \textbf{Out-of-vocabulary error}: when extracting keywords, we also need to carefully handle the out-of-vocabulary error, where some phases were not included during pretraining or aligning phases for the \textsc{parser-LLM} --- such issue could significantly compromise the results of keyword extraction. For instance, decomposing unfamiliar proper nouns, such as company names that a large language model (LLM) has not encountered—into multiple keywords can lead to the retrieval of irrelevant information. This issue arises because the model lacks prior knowledge of these terms, making it challenging to interpret and process them accurately. 
 \end{itemize}

\vspace{0.15em}
\textbf{Instruction Set for \textsc{parser-LLM}.} We construct a multi-task instruction set to train our \textsc{parser-LLM} to enable it to perform both retrieval determination and keyword extraction tasks in a single inference pass. These tasks are guided by tailored instructions designed to address the specific challenges outlined above. We utilize \textsc{GPT-4o}, one of the most advanced LLM API available, to generate high-quality ground truths for these tasks, ensuring the instruction set is comprehensive and robust. To fulfill the requirements listed above, we integrate the following implementation: 

\begin{itemize}[topsep=5pt, leftmargin=1em]
\item \textbf{Support multi-instruction following}: to improve efficiency, the \textsc{parser-LLM} is trained to generate both the retrieval determination and the extracted keywords simultaneously within a single pass at the inference time. This requires the model to effectively interpret the interdependence of these tasks. Since 6B scale LLMs often exhibit weaker multi-task instruction-following capabilities, we carefully designed and manually tuned an instruction template that enables our smaller Parser-LLM to achieve acceptable zero-shot performance. 

\vspace{0.15em}
\item \textbf{Time information integration}: recognizing that keyword extraction effectiveness can vary with time sensitivity, we specifically design instructions that guide the \textsc{parser-LLM} to handle both positive and negative temporal dependencies in user queries. The data in the instruction set for these cases is generated by \textsc{GPT-4o}, ensuring accurate modeling of scenarios where temporal details either enhance or hinder retrieval effectiveness. This approach enables the model to dynamically adjust its handling of time-specific information based on the context of the query.

\vspace{0.15em}
\item \textbf{Enhancing with multi-lingual search}: to address the limitations of monolingual keyword extraction, the instruction set explicitly incorporates multi-language instructions. This enables the \textsc{parser-LLM} to extract keywords from high-context and deductive queries expressed in different languages. Training data includes examples emphasizing the ambiguity and richness of languages like Chinese, ensuring the model effectively captures essential keywords across linguistic boundaries. This multilingual approach reduces the risk of misinterpretation and improves retrieval accuracy.

\vspace{0.15em}
\item \textbf{Resolving out-of-vocabulary errors}: out-of-vocabulary errors are mitigated through the construction of synthetic data containing made-up words designed to simulate unfamiliar terms, such as novel proper nouns or company names. By learning from these examples, the \textsc{parser-LLM} becomes more adept at handling previously unseen vocabulary without decomposing such terms into irrelevant keywords. This approach ensures the robustness of the keyword extraction process, even in scenarios involving rare or new terms.
\end{itemize}
\vspace{-0.5em}

\vspace{0.15em}
\textbf{Training Details.}
We choose the \textsc{Yi-6B-Chat} model as the initial model to obtain the \textsc{parser-LLM} throughput instruction fine-tuning. Note that compared with the \textsc{Yi-6B-Base}, \textsc{Yi-6B-Chat} has been supervised fine-tuned to obtain basic instruction-following capabilities. 
We use the prepared instruction set notated as $\mathcal{I}$ to train the base model in a standard auto-regressive manner, with the optimization objective defined as follows:
\vspace{-1em}

\begin{equation*}
\mathcal{L} = min_{\theta} \ \mathbb{E}_{ j \sim \mathcal{I}} \left[ - \log p( \Omega_j \mid \texttt{Request}_j; \theta) \right]
\end{equation*}

\noindent where $\mathcal{L}$ represents the negative likelihood loss function that needs to be minimized, $\Omega_j$ is the expected output of the multi-task instruction for the j-th  $\texttt{Request}_j$, $\theta$ denotes the model parameters. The expectation $\mathbb{E}_{ j \sim \mathcal{I}}$ is calculated by sampling through the instruction set $\mathcal{I}$.

We train the model for three epochs, with a maximum sequence length of 16K tokens. The learning rate is set to $5 \times 10^{-6}$ initially, and a cosine learning rate scheduler was employed, and the mininal learning rate is set to $1.5 \times 10^{-6}$. To prevent overfitting, the training data is shuffled at the beginning of each epoch. We the FusedAdam optimizer with the hyper-parameter configured as $\beta_1 = 0.9$, $\beta_2 = 0.95$, and $\epsilon = 1 \times 10^{-8}$. The global batch size is set to $16$. The training computation was conducted on an NVIDIA server equipped with four  H800 GPUs by the DeepSpeed~\cite{deepspeed} parallel training framework.

\subsection{Search Result Re-ranking}
\label{sec:ranker}

To eliminate the bias introduced by search engine API, considering the irrelevant factors such as click-through rate and dwell time, we leverage open-source the \textsc{BGE-M3} model~\cite{chen2024bge} to assess the relevance of search results based on the extracted keywords and generate the ranking score.

We introduce a new \textit{mixed ranking strategy}, where the \textsc{BGE-M3} model can compute relevance scores for each retrieved item in the reranking pool, which consists of two levels of granularity derived from the HTML file: the snippet and the full content. We consider both the snippet and the full content because they can provide different information for the reranking model: the snippet, as a concise summary provided by the search engine, offers high-level relevance but may lack detailed accuracy; while the full content, extracted from the HTML body, contains comprehensive details but can obscure high-level relevance due to information overload.
In our \textit{mixed ranking strategy}, each snippet and the corresponding full content are treated as an independent item, which doubles the number of candidates to $2n$ for $n$ HTML files. This approach ensures that both high-level and detailed relevance are considered in the ranking process. The reranked results are then aggregated, selecting the top-K distinct HTML files based on the highest ranking achieved by either their snippet or full content. Note that the results retrieved from auxiliary keyword lists (e.g., from multilingual queries) are combined with those from the main keyword lists, broadening the candidate pool and enhancing coverage. This strategy ensures that the final selection is optimized for both granularity types, balancing high-level relevance and detailed accuracy.



\begin{figure*}[htbp]
\centering
\centerline{\includegraphics[width=0.95\linewidth]{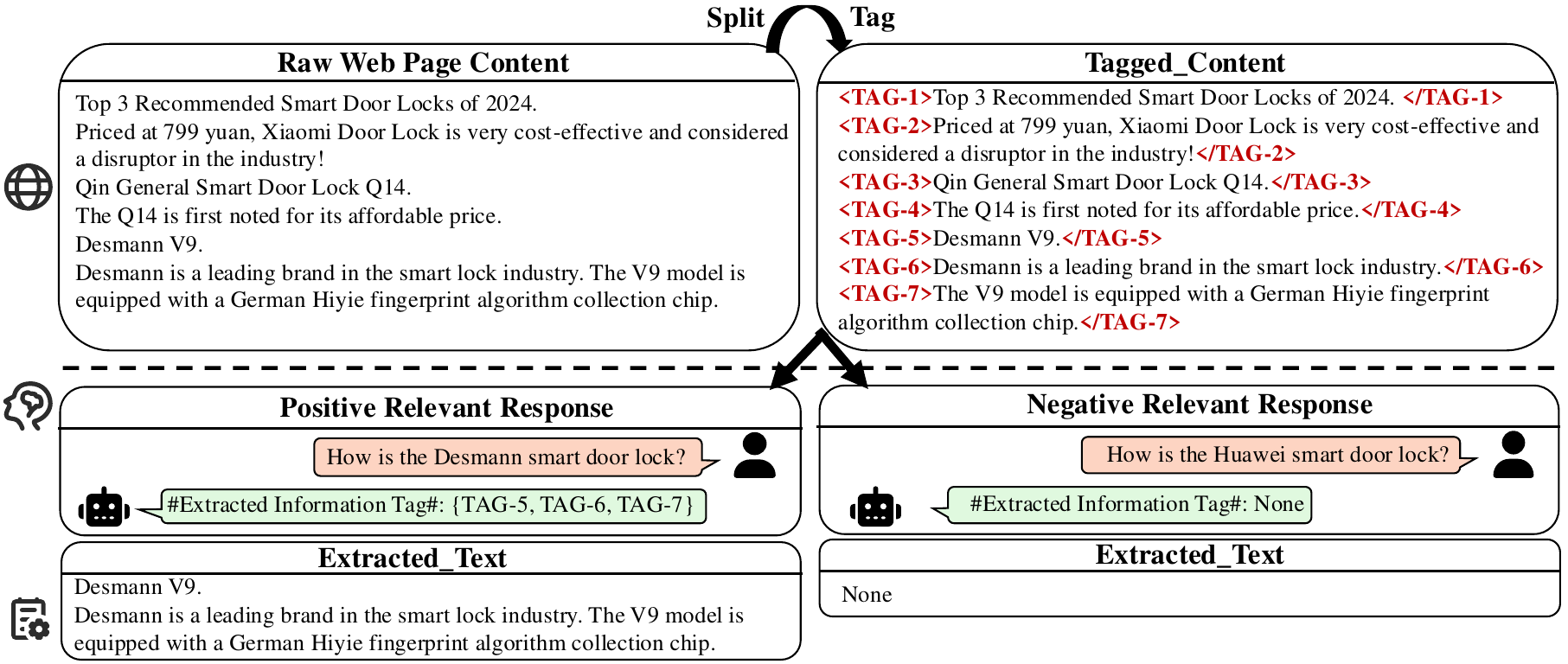}}
\caption{An illustration about pre-processing the retrieved HTML file and Extractor LLM processing procedure.}
\label{fig:case1}
\vspace{-1.5em}
\end{figure*}

\section{EXTRACT RELEVANT INFORMATION}
\label{sec:train}


As we introduced in Section \ref{sec:isag}, the second core technique functionality in our Internet search augmented generation paradigm is the \textsc{extractor-LLM}, which should be able to accurately extract relevant information without introducing significant computational overhead at the inference time. Toward this end, in Section \ref{sec:preprocessing}, we first enumerate a preprocessing mechanism for the retrieved HTML file to generate tags for the content so that the \textsc{extractor-LLM} only outputs the corresponding tags of the content to save the budget of generated tokens.  Then we introduce a novel two-phase training strategy to train the \textsc{extractor-LLM}, where the primary goal is to fulfill the functionality of determining if the relevant content exists or not and extracting tags of the relevant and user-preferred texts from the pre-processed contents. We summarize this two-phase training paradigm below:

\begin{itemize}[topsep=5pt, leftmargin=1em]
    \item \textbf{SFT phase}: the first phase of training uses supervised fine-tuning (SFT) to train the \textsc{extractor-LLM}'s basic ability to extract relevant information, where the construction of instruction set is explained in Section \ref{sec:instruct} and the SFT training details are in Section \ref{sec:traindetail}. 
    \item \textbf{DPO phase}: the second phase further aligns the extraction with the preference of a judge after training with reinforcement learning technique, i.e., direct preference optimization (DPO)\cite{rafailov2024direct}, where the data preparation is introduced in Section \ref{sec:dataprep} and DPO training details are in Section \ref{sec:dpodetail}. 
\end{itemize}

\vspace{-0.75em}
\subsection{Preprocessing the HTML File} 
\label{sec:preprocessing}

Our pre-processing mechanism applies a heuristic to generate splitting tags for the HTML body content. This heuristic, implemented as a finite state machine (FSM), uses special punctuation marks such as \texttt{.}', \texttt{!}', and `\texttt{?}' as delimiters to segment the content, enclosing each segment within corresponding tags. For example, the ${i}$-th segment is enclosed within the tags \texttt{<TAG-${i}$>} and \texttt{</TAG-${i}$>}. Note that the finite state machine is also in charge of handling some relatively complex situations; for example, when the period `\texttt{.}' is processed, the finite state machine manages to distinguish if the `\texttt{.}' indicates the end of a sentence or a decimal point based on the status whether `\texttt{.}' follows numerical values. 
Figure \ref{fig:case1}-(top) shows a concrete example of this pre-processing procedure: The retrieved context from the HTML file introduces three types of smart door locks: Xiaomi, Qin General, and Desmann; our finite state machine is able to split the contents with different tags; for example, the introduction for Xiaomi, Qin General, and Desmann is enclosed with $\{$\texttt{TAG-2}$\}$, $\{$\texttt{TAG-3},\texttt{TAG-4}$\}$, $\{$\texttt{TAG-5}, \texttt{TAG-6},\texttt{TAG-7}$\}$, respectively.

During the inference time, once the content is segmented, the \textsc{extractor-LLM} performs two main tasks: (i) determining if the content includes information relevant to the input requests; and (ii) extracting the relevant tags and their content. Formally, the \textsc{extractor-LLM} takes the \texttt{Request} and each of the tagged content from Internet search API notated as \texttt{Tagged\_Content} as inputs. If there is relevant information in the content, the \textsc{extractor-LLM} outputs the corresponding tags; otherwise, the \textsc{extractor-LLM} outputs \texttt{None} to indicate there is no relevant information in this content. 
Figure \ref{fig:case1}-(bottom) shows how the Extractor LLM determines the output for the same  \texttt{Tagged\_Content} with different generative inference \texttt{Request}s: On the left-hand side, the inference \texttt{Request} demands the information about Desmann smart door lock, so the Extractor LLM outputs tags \{\texttt{TAG-5},\texttt{TAG-6},\texttt{TAG-7}$\}$; On the right-hand side, the inference \texttt{Request} demands the information about Huawei smart door lock, so the Extractor LLM returns \texttt{None} since there is no relevant information about Huawei smart door lock from this \texttt{Tagged\_Content}.

\subsection{Instruction Set Construction for SFT}
\label{sec:instruct}

In the first SFT phase, we aim to obtain a fine-tuned model that can explicitly determine the existence of the relevant information and the corresponding tags (if they exist). For this purpose, the central goal of our instruction set construction is to collect training data that can mirror the process within the Internet search augmented generation paradigm. We find that the essential requests for the constructed instruction set lie in three aspects: 

\begin{itemize}[topsep=5pt, leftmargin=1em]
 \item \textbf{Diversity of the requests and contents}: since we expect the Internet search augmented generation paradigm to be generally applicable to any generative inference requests in an API serving system, the \textsc{extractor-LLM} should be able to handle various information-extracting scenarios. Thus, the instruction set should include a collection of $<$\texttt{Request}, \texttt{Tagged\_Content}$>$ data pairs from diversified real-world scenarios.  

 \vspace{0.15em}
 \item \textbf{Necessary reasoning procedure}: during our preliminary exploration, we found that the \textsc{extractor-LLM} would generalize poorly if we directly fed the desired \texttt{TAG}s during the instruction fine-tuning phase without providing additional information. We speculate that this is because some necessary reasoning procedure is missing in this instruction data so that the fine-tuning procedure fails to capture the implicit relation between the \texttt{Request}, the \texttt{Tagged\_Content}, and the desired output \texttt{TAG}s. This indicates that some reasoning information about the extracting procedure will be necessary for instruction fine-tuning.      

\vspace{0.15em}
 \item \textbf{Accurate extraction with minimal noise}: when preparing the instruction set, the common practice involves calling existing LLM APIs to significantly automate and accelerate this labor-intensive process. However, such an approach comes at the cost of degraded instruction data quality, as the generated content can be noisy, containing irrelevant, inaccurate, or misleading information. Thus, it is necessary to apply some additional refinement mechanisms to ensure the final instruction set is both accurate and clear.  
\end{itemize}

In order to satisfy these requests, we enumerate how we construct the instruction set as below:

\vspace{0.15em}

\textbf{Data Source.} To ensure quality and diversity in our instruction set, we collect real-world production data from 01AI's inference API serving system. Concretely, we collect raw API requests during one week. Given that each request is associated with multiple web pages returned by the search engine, to avoid excessive repetition and maintain request diversity, we randomly sampled only two contents from all the returned web pages for each request. 

\vspace{0.15em}

\textbf{Prepare Instruction Data with LLMs.} After collecting the raw data from the production environment, we need to use LLMs to process the input data by the format of $<$\texttt{Request}, \texttt{Tagged\_Content}$>$, to automate the extraction process of the desired collection of \texttt{TAG}s. To introduce some information for the necessary reasoning procedure, we design the following procedure: we ask the LLM not only to generate the desired collection of \texttt{TAG}s, but also to generate a short \texttt{Summary} of the corresponding content associated with each of the \texttt{TAG}s. Formally, this LLM-aided instruction construction procedure can be defined as follows:

\vspace{-0.75em}

\begin{equation*}
\begin{aligned}
    &<\texttt{Request}, \texttt{Tagged\_Content}> \\
\rightarrow &<\texttt{Request}, \texttt{Tagged\_Content}, \{<\texttt{Summary}_{k_i}, \texttt{TAG}_{k_i}>_{i=1,...,n} \}> 
\end{aligned}
\end{equation*}

\noindent which means, for each ${<}\texttt{Request}, \texttt{Tagged\_Content}{>}$ pair, we generate a set of $n$ relevant ${<}\texttt{Summary}_{k_i},\texttt{TAG}_{k_i}{>}$ pairs. 
Besides such instructions notated as \texttt{Instruction-A}, we find that empirically it is also necessary for the instruction set to include the instructions that only contain the \texttt{TAG}s without the \texttt{Summary}. To obtain such instruction data, for each \texttt{Instruction-A}, we manually remove the \texttt{Summary} to generate \texttt{Instruction-B}, which can be formalized as:
\vspace{-0.25em}

\begin{equation*}
\begin{aligned}
    &<\texttt{Request}, \texttt{Tagged\_Content}, \{<\texttt{Summary}_{k_i}, \texttt{TAG}_{k_i}>_{i=1,...,n} \}> \\
\rightarrow &<\texttt{Request}, \texttt{Tagged\_Content}, \{ <\texttt{TAG}_{k_i}>_{i=1,...,n} \}> 
\end{aligned}
\end{equation*}

\noindent During SFT, we include both \texttt{Instruction-A} and \texttt{Instruction-B}. Empirically, this mixture of two instruction types works well --- we speculate that this is owing to the \texttt{Instruction-A} helps to bridge the semantic meaning of the context ($<$\texttt{Request}, \texttt{Tagged\_Content}$>$) to the associated \texttt{TAG}s, while \texttt{Instruction-B} further help to teach the \textsc{extractor-LLM} to follow the desired information extracting instruction under the Internet search augmented generation paradigm.

\vspace{0.15em}
\textbf{Improve the Robustness of the Instruction Set.} We also improve the robustness of the instruction set before, during, and after the processing by the OpenAI API calls. 

\begin{itemize}[topsep=5pt, leftmargin=1em]
\item \textbf{Before} processing by the API, we carefully filter the input pair by the format of ${<}\texttt{Request}, \texttt{Tagged\_Content}{>}$, and remove the pairs with \texttt{Tagged\_Content} less than 100 tokens --- such short web page from the search engine API usually contained limited helpful information. 

\vspace{0.15em}
\item \textbf{During} processing by the API, since the API does not generate deterministic output in multiple inferences given the same input prompt, for each ${<}\texttt{Request}, \texttt{Tagged\_Content}{>}$ pair, we process the same pair twice with different random seeds; if the results are consistent, we keep the instruction otherwise we abandon such contradictory result. We find such a policy can remove a lot of low-quality data and decrease the uncertainty and noise introduced by API calls. 

\vspace{0.15em}
\item \textbf{After} processing by API, since it is important for the \textsc{extractor-LLM} to determine whether there is relevant information from the content, we manually check the ratio of the instructions that should return \texttt{None} as the results, and ensure the ratio of \texttt{None} to be $5\%$ in the instruction set, which is determined based on empirical tests. 
\end{itemize}

\subsection{SFT Training Details}
\label{sec:traindetail}

We choose the \textsc{Yi-9B-Chat-16K} model as the initial model for instruction fine-tuning. Note that compared with \textsc{Yi-9B-Base-16K}, this model has been supervised and fine-tuned to obtain basic instruction-following capabilities. 
We use the prepared instruction set notated as $\mathcal{I}$ to train the base model in a standard auto-regressive manner, with the optimization objective defined as follows:

\begin{equation*}
\mathcal{L} = min_{\theta} \ \mathbb{E}_{ j \sim \mathcal{I}} \left[ - \log p( \Omega_j \mid <\texttt{Request}_j, \texttt{Tagged\_Content}_j>; \theta) \right]
\end{equation*}

\noindent where $\mathcal{L}$ represents the negative likelihood to be minimized, $\Omega_j$ is the expected output for the j-th ${<}\texttt{Request}_j,\\ \texttt{Tagged\_Content}_j{>}$ pair, $\theta$ denotes the model parameters. The expectation $\mathbb{E}_{ j \sim \mathcal{I}}$ is taken over the instruction set $\mathcal{I}$. We train the model for three epochs, with a maximum sequence length of 16K tokens. The learning rate is set to $5 \times 10^{-6}$ initially, and a cosine learning rate scheduler was employed. To prevent overfitting, the training data is shuffled at the beginning of each epoch. We choose the FusedAdam optimizer with the hyper-parameter configured as $\beta_1 = 0.9$, $\beta_2 = 0.95$, and $\epsilon = 1 \times 10^{-8}$. The global batch size is set to $16$. The training computation was conducted on four NVIDIA H800 GPUs, each with 80GB of memory by the DeepSpeed~\cite{deepspeed} parallel training system.

\begin{figure}[htbp]
\centering
\centerline{\includegraphics[width=0.85\linewidth, height=0.3\textheight]{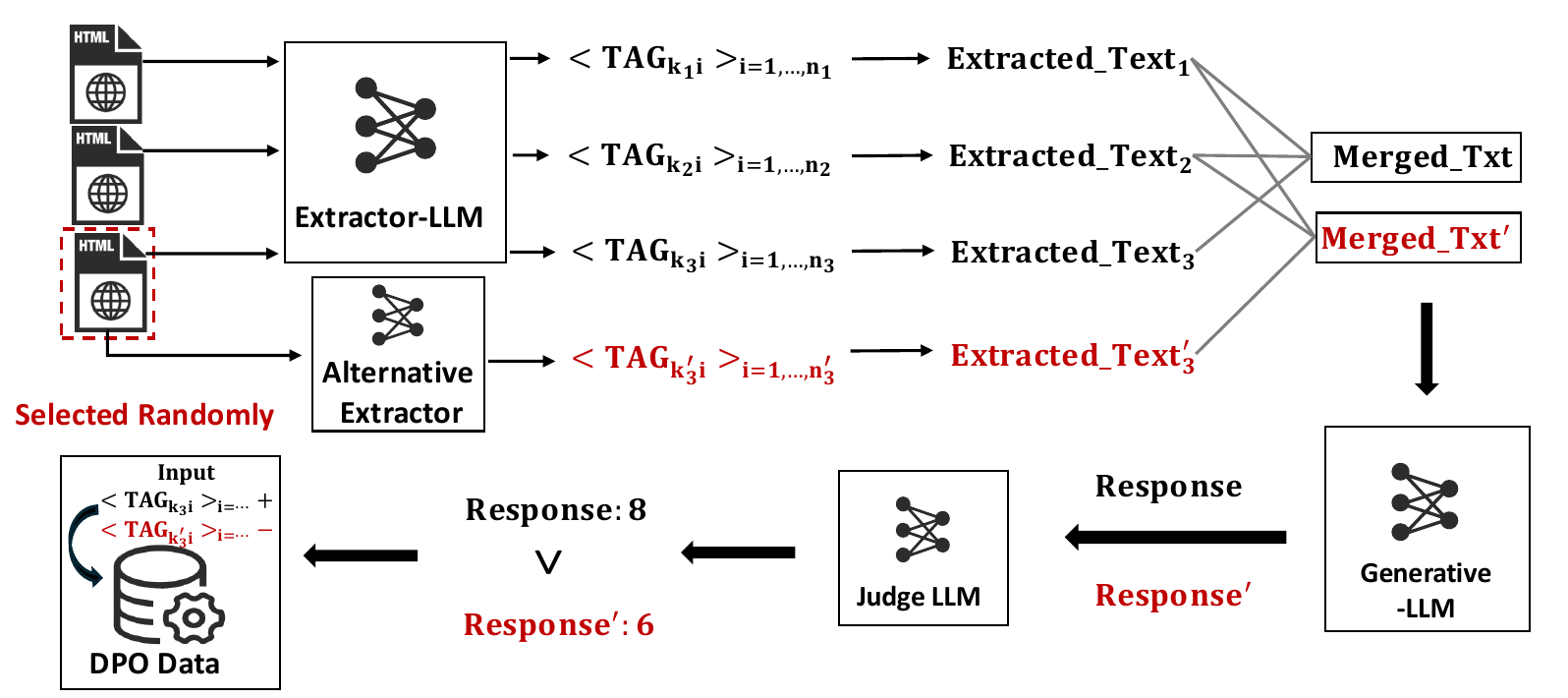}}
\caption{An illustration Of DPO Training Preparation.}
\label{fig:DPO}
\vspace{-1.5em}
\end{figure}

\vspace{-0.75em}
\subsection{Data Preparation for DPO}
\label{sec:dataprep}
In the second DPO phase, we further optimize the \textsc{extractor-LLM}'s output TAGs towards an end-to-end preference estimated by OpenAI API (i.e., \textsc{GPT-4o}) considered as a \textit{LLM judge}. Note that we can also replace this role with human preference. We modify the single answer grading prompt provided in a prior study~\cite{zheng2024judging} to design the judging instruction. To be more specific, we add more detailed grading explanations and standards while asking the Judge LLM to be more critical and precise when scoring.
We denote \textsc{extractor-LLM} after SFT as $\pi_{ref}$. The data preparation for the DPO phase needs to provide a set of data in the format of $\{\texttt{Request}, \texttt{Tagged\_Content}, \texttt{y}^+, \texttt{y}^-\}$, where $\texttt{y}^+$ indicates the extracted TAGs is preferred by the LLM judge, and $\texttt{y}^-$ indicates they are less favored. We explain how we construct $\texttt{y}+$ and $\texttt{y}-$ from the \texttt{TAG}s extracted by different extractors as below:

\begin{itemize}[topsep=5pt, leftmargin=1em]
 \item  \textbf{Step 1. Generate response using} {$\pi_{\text{ref}}$} \textbf{as extractor}: given a \texttt{Request}, for each \texttt{Tagged\_Content} retrieved by the Internet search engines, we firstly use our \textsc{extractor-LLM} after SFT noted as $\pi_{\text{ref}}$ to perform extractions and obtain ${<}\texttt{TAG}_{k_i}{>}_{i=1,...,n}$. The extracted ${<}\texttt{TAG}_{k_i}{>}_{i=1,...,n}$ are referred back to the $m$ corresponding texts, obtaining \texttt{Extracted\_Text}. Then we ensemble all the extracted text ${<}\texttt{Extracted\_Text}_{k}{>}_{k=1,...,m}$ from the corresponding tagged contents ${<}\texttt{Tagged\_Content}_{k}{>}_{k=1,...,m}$  into \texttt{Merged\_Text}. Lastly, we combine the \texttt{Merged\_Text} and \texttt{Request} together as an enriched prompt, which is sent to the generative LLM to generate a \texttt{Reponse}.

\vspace{0.15em}
\item \textbf{Step 2. Permute response by partial replacement}: we use \textsc{GPT-4o} as the alternative extractor model to perform \texttt{TAG} extraction, enabling the judge to make comparisons during the later evaluation process.
To implement the replacement, we randomly select one 
tagged content, denoted as $\texttt{Tagged\_Content}_{i}$, from all the tagged contents returned by Internet search engines. For this $\texttt{Tagged\_Content}_{i}$, we replace its corresponding $\texttt{Extracted\_Text}_{i}$ with $\texttt{Extracted\_Text}_{i}^{'}$, where $\texttt{Extracted\_Text}^{'}_{i}$ is the text extracted by \textsc{GPT-4o} from $\texttt{Tagged\_Content}_{i}$ using the same procedure described in \textbf{Step 1}. All other extracted texts, denoted as ${<}\texttt{Extracted\_Text}_{k}{>}_{k=1,...,i-1,i+1,...,m}$, remain unchanged, and the updated $\texttt{Extracted\_Text}_{i}'$ is then merged with all the other unchanged texts ${<}\texttt{Extracted\_Text}_{k}{>}_{k=1,...,i-1,i+1,...,m}$ to construct $\texttt{Merged\_Text}'$. This $\texttt{Merged\_Text}'$ is combined with $\texttt{Request}$ to form an enriched prompt and later be fed into the backbone \textsc{generative-LLM} to generate a new response, $\texttt{Reponse}^{'}$. The upper part of Figure \ref{fig:DPO} illustrates the process of generating both $\texttt{Reponse}$ and $\texttt{Reponse}^{'}$ for a given \texttt{Request} and selected $\texttt{Tagged\_Content}_{i}$.

\vspace{0.15em}
\item  \textbf{Step 3. Obtain the preference from the LLM judge}: for each given \texttt{Request} and a chosen $\texttt{Tagged\_Content}_{i}$, we now have two different responses generated by the backbone \textsc{generative-LLM} (i.e., \textsc{Yi-lightning}). \texttt{Reponse} is generated with information extracted entirely by $\pi_{\text{ref}}$. And $\texttt{Reponse}^{'}$ uses the majority of extractions by $\pi_{\text{ref}}$, except for the chosen $\texttt{Tagged\_Content}_{i}$, where the extraction is performed by \textsc{GPT-4o}. We ask the LLM judge to score each response separately based on how well it answers the \texttt{Request}. The better-scored response is chosen as the preferred response, denoted by $\texttt{Response}^{+}$, and the other is denoted by $\texttt{Response}^{-}$.

\vspace{0.15em}
\item  \textbf{Step 4. Construct the dataset}: up to this point, we have a pair of responses, $\texttt{Response}^{+}$ and $\texttt{Response}^{-}$, for a given \texttt{Request} and a chosen $\texttt{Tagged\_Content}_{i}$. For each response, we trace back to the corresponding tags noted as ${<}\text{TAG}_{k_i}{>}_{i=1,...,n}$ or ${<}\text{TAG}_{k'_i}{>}_{i=1,...,n'}$ extracted from $\texttt{Tagged\_Content}_{i}$, depending on which is used to generate this response. And the extracted \texttt{TAG}s corresponds to $\texttt{Response}^{+}$ is constructed as $\texttt{y}^+$, while the extracted $\texttt{TAG}$s corresponds to $\texttt{Response}^{-}$ is $\texttt{y}^-$. We then construct $\{\texttt{Request}, $\texttt{Tagged\_Content}$, \texttt{y}^+, \texttt{y}^-\}$ as a piece of data in the DPO dataset. For each $\texttt{Request}$, we repeat \textbf{Step 2} and \textbf{Step 3} only twice to obtain two such pieces of data, so that more \texttt{Request}s can be included in the dataset. This can increase the diversity of different $\texttt{Request}$s during the DPO training. 

\end{itemize}

\vspace{-0.75em}
\subsection{DPO Training Details}
\label{sec:dpodetail}

Given the training dataset prepared for DPO, we want to obtain a trainable actor model denoted as $\pi_{\theta}$ for which it can extract \texttt{TAG}s more aligned with the judge's preference. We initialize $\pi_{\theta}$ with the reference model $\pi_{\text{ref}}$, which is the fine-tuned \textsc{extractor-LLM} after SFT phase. We use the prepared DPO training dataset noted as $\mathcal{D}$ to train the actor model $\pi_{\theta}$, with the optimization objective defined as follows:
\vspace{-1em}

\begin{equation*}
\mathcal{L} = E_{(x,\texttt{y}^+,\texttt{y}^-)\sim D}[-\log\sigma(\beta log \frac{\pi_{\theta}(\texttt{y}_{+}|x)}{\pi_{ref}(\texttt{y}_{+}|x)} - \beta log \frac{\pi_{\theta}(\texttt{y}_{-}|x)}{\pi_{ref}(\texttt{y}_{-}|x)})]
\end{equation*}

\vspace{-0.25em}
\noindent where $x$ consist of a given \texttt{Request} and a chosen $\texttt{Tagged\_Content}$, $\pi_{\theta}(\texttt{y}_{+}|x)$ is the probability of predicting $\texttt{y}^+$ when giving $x$ as input, and $\pi_{\theta}(\texttt{y}_{-}|x)$ is the probability of predicting $\texttt{y}^-$ when giving $x$ as input. $\pi_{ref}(\texttt{y}_{+}|x)$ and $\pi_{ref}(\texttt{y}_{-}|x)$ are defined in a similarly way. $\sigma$ is the sigmoid function, and $\beta$ is the coefficient hyperparameter~\cite{rafailov2024directpreferenceoptimizationlanguage}.

Minimizing $\mathcal{L}$ suggests that we expect $\pi_{\theta}$ to extract \texttt{TAG}s aligned with $y^+$ and less like $y^-$ when given $x$. Meanwhile, we don't want $\pi_{\theta}$ to deviate too much from $\pi_{ref}$, so that it can still extract properly, and doesn't turn into something that is rated very high, but has a garbled extraction of \texttt{TAG}s.



\section{Evaluation}
\label{sec:eval}

Our empirical study lies in two aspects: (\underline{i}) the end-to-end performance of the Internet search augmented generation paradigm; and (\underline{ii}) the capability of the \textsc{extractor-LLM} to identify the relevant information. Concretely, we aim to answer the following four questions:

\begin{itemize}[topsep=5pt, leftmargin=1em]
    \item What is the end-to-end performance of the response quality when comparing the proposed Internet search augmented generation paradigm to other RAG paradigms?
    \item What is the generative cost in terms of the processed input tokens when contrasting the proposed Internet search-augmented generation paradigm with other RAG approaches?

    \item Can the \textsc{extractor-LLM} extract the relevant information accurately and precisely?
    \item How robust is the \textsc{extractor-LLM} that can reject inappropriate contents rather than generate hallucinations when there is no relevant information?
\end{itemize}

\subsection{End-to-End Evaluation} 

To evaluate the (\underline{i}) end-to-end performance regarding the quality of the generated content and (\underline{ii}) the generative cost in terms of the processed input tokens between the proposed Internet search augmented generation paradigm and other RAG paradigms, we design the following configurations:

\vspace{0.15em}
\textbf{Evaluation Task Data.} The evaluation dataset is constructed from real-world data. We randomly sampled $500$ queries from 01AI's real-world API request --- we confirmed that there is no overlap between this evaluation dataset and the raw data we used to prepare the instruction set introduced in Section \ref{sec:instruct}. After deduplication --- $463$ unique requests remained, accompanied by the content scraped from corresponding web pages returned by search engines. 

\vspace{0.15em}
\textbf{Baseline Systems.} We use \textit{Internet-SAG-Ext} to denote our proposed Internet search augmented generation paradigm. 
We consider two baseline systems: 

\begin{itemize}[topsep=5pt, leftmargin=1em]
\item \textbf{Internet-SAG-Naive:} in this baseline system, we remove the \textsc{Extractor-LLM} component in \textit{Internet-SAG-Ext}, and rely on the \textsc{Generative-LLM}'s ability to comprehensively integrate the information from the search API returned contents. Concretely, all of the content retrieved from the Internet via search engines is simply concatenated and directly used as reference materials for the \textsc{Generative-LLM} to respond to the API request.

\item \textbf{VectorDB-RAG:} we configure a vector database to manage all of the Internet-retrieved content in the evaluation task data. Concretely, we employ the standard RAG paradigm. 
During the building phase, all content associated with the API request is split into overlapping chunks according to specified chunk size and chunk overlap parameters. The embedding model is then used to generate vector representations of each chunk, which are stored in the vector database. During the retrieval phase, a vector representation of the request is generated, and the top-k most relevant chunks (based on cosine similarity) are retrieved; 
The top-k relevant chunk is concatenated to form the final reference materials for the \textsc{Generative-LLM}. We use BGE-m3~\cite{chen2024bge} from BAAI as the embedding model, FAISS~\cite{douze2024faiss} as the vector store, with hyperparameters set as: chunk-size = 512, chunk-overlap = 128, top-k = 12, neighbor-num = 1 (indicating that chunks immediately before and after the relevant chunk should be included, if available).

\end{itemize}

\vspace{0.15em}
\textbf{\textsc{Generative-LLM} Configuration.} We evaluate three models with different scales as the \textsc{Generative-LLM}s in different RAG systems, including the open-source models \textsc{Qwen2-72B} and \textsc{Yi-34B}, as well as the closed-source model \textsc{Yi-large}. 

\vspace{0.15em}
\textbf{Evaluation Metrics.}
We employ \textsc{GPT-4o} as the judge to evaluate the quality of responses generated by different RAG systems. To maintain consistency, we specifically use the \textsc{GPT-4o-2024-05-13} version for this evaluation.
To address positional bias when applying \textit{LLM judge}, we switch the order of the two responses specified in the \texttt{prompt-JUDGE} and conduct two separate evaluations~\cite{zheng2024judging}. Responses that yield consistent results across both evaluations are accepted, while those with inconsistent outcomes are classified into a special positional bias (P-BIA) category. Regarding length and structure biases, the generation prompt \texttt{prompt-GEN} explicitly instructs the model to provide detailed answers and use markdown formatting unless otherwise specified in the API request. For the judgment prompt \texttt{prompt-JUDGE}, used by GPT-4o for pairwise comparison, we reuse the prompt provided in a prior study~\cite{zheng2024judging}.

\vspace{0.15em}
\textbf{Experiment Results.}
The experimental results of end-to-end performance regarding the quality of the generated response are summarized in Table~\ref{tbl:test}, where each row presents the performance outcomes (WIN, TIE, LOSE, and P-BIA) of the \textit{Internet-SAG-Ext} paradigm compared to baseline paradigm equipped with different \textsc{Generative-LLM}s. 
On the other hand, the generative cost in terms of the consumed input tokens by the \textsc{Generative-LLM} is shown in Table \ref{tbl:input_token_cost}.

\begin{table}[h]
	\centering
	\caption{End-to-End evaluation on Internet-SAG-Ext system versus other two baseline systems, includes \textit{Internet-SAG-Naive} and \textit{VectorDB-RAG}, employed with various GEN-LLMs (\textsc{Generative-LLM}s). The value in each cell denotes the number of WIN, TIE, LOSE, or P-BIA respectively.}
	\begin{tabular}{lccccc}
	\toprule 
	Baseline SYS & \textsc{Gen-LLM} & WIN & TIE & LOSE & P-BIA\\
	\midrule
	\multirow{3}{*}{Internet-SAG-Naive} & \textsc{Yi-large} & 143 & 22 & 89 & 209\\
     & \textsc{Qwen2-72B} & 134 & 20 & 93 & 216\\
	 & \textsc{Yi-34B} & 137 & 22 & 85 & 219\\
    \cmidrule(lr){2-6}
	\multirow{3}{*}{VectorDB-RAG} & \textsc{Yi-large} & 134 & 15 & 95 & 219\\
     & \textsc{Qwen2-72B} & 258 & 23 & 66 & 116\\
	 & \textsc{Yi-34B} & 185 & 12 & 61 & 205\\
	\bottomrule
	\end{tabular}
	\label{tbl:test}
\end{table}

\begin{table}[h]
    \centering
    \caption{The total input token cost of generative LLM provided by different systems. M means million. The second column indicates the number of total input token cost.}
    \begin{tabular}{c c}
    \toprule
    SYS & \#Input Token \\ 
    \midrule
    Internet-SAG-Ext & 2.04 M \\
    Internet-SAG-Naive & 3.88 M \\
    VectorDB-RAG & 2.59 M \\
    \bottomrule
    \end{tabular}
    \label{tbl:input_token_cost}
    \vspace{-1.5em}
\end{table}

\vspace{0.15em}
\textbf{Discussion.}
The experimental results indicate that our proposed \textit{Internet-SAG-Ext} paradigm consistently outperforms the baseline systems across different \textsc{Generative-LLM}s and achieves the lowest cost in terms of the input tokens processed by \textsc{Generative-LLM}s.
When using the \textsc{Yi-large} as \textsc{Gererative-LLM}, \textit{VectorDB-RAG} demonstrates a slight advantage over \textit{Internet-SAG-Naive}. This suggests that simply concatenating content returned by the search engine and passing it to the generative model is insufficient. A plausible explanation is that occupying an excessively long sequence window can degrade model performance, as noted in~\cite{hsieh2024ruler} concerning inflated window length.
On the other hand, when using the \textsc{Qwen2-72B}, \textit{VectorDB-RAG} performs worse than \textit{Internet-SAG-Ext}. In many responses generated by \textsc{Qwen2-72B} within the \textit{VectorDB-RAG} paradigm, we observed that the model autonomously created its own questions based on the references and then answered them, completely ignoring the original request specified in the \texttt{prompt-GEN}. This issue may arise from the pre-configured chunk size and top-K settings in \textit{VectorDB-RAG}, which could have introduced irrelevant information during retrieval. When the proportion of irrelevant information becomes too large, it may trigger such observed issues in \textsc{Qwen2-72B}.
When using the \textsc{Yi-34B} model, \textit{Internet-SAG-Naive} outperforms \textit{VectorDB-RAG}. As mentioned earlier, the inherent issue with \textit{VectorDB-RAG} lies in its pre-configured chunk size and top-K settings, which may introduce irrelevant information during retrieval. Although we attempted to filter out irrelevant information using instructions within \texttt{prompt-GEN}, the other two models, except for \textsc{Yi-large}, exhibited lower resistance to interference and reduced robustness when handling irrelevant information.
Nevertheless, \textit{Internet-SAG-Ext} demonstrates superior performance across various scenarios. After the \textsc{Extractor-LLM} performs information extraction, the text length is reduced, preventing the model from occupying too much of the sequence window and degrading its performance. Additionally, irrelevant information is filtered out, reducing the risk of the model being biased by irrelevant data.
Moreover, since most LLM service platforms charge based on the number of tokens processed, the information processed by \textit{Internet-SAG-Ext} requires fewer input tokens for the generative LLM compared to the baselines, providing a cost advantage as well. Specifically, we reduce $21\%$ and $47\%$ input token cost for Generative LLM compared to \textit{VectorDB-RAG} and \textit{Internet-SAG-Naive} respectively.

\subsection{\textsc{Extractor-LLM} Evaluation}
In this section, we comprehensively assess the capabilities of our \textsc{Extractor-LLM}, primarily focusing on extracting relevant information, accurately locating tags, and effectively rejecting irrelevant content.

\vspace{0.15em}
\textbf{Benchmark. }We discuss three benchmarks to comprehensively evaluate the context retrieval ability of our model in different aspects, including the synthetic, open-source, and real-world benchmarks. We begin by adopting a standard methodology in the retrieval field to construct our \textit{Synthetic Benchmark}, which enables fine-grained evaluations through controlled variations in context length and the depth of the answer where it is inserted in the context. The \textit{Open-source Benchmark} adapts established datasets to assess specific retrieval capabilities, providing extensive test cases. The \textit{Real-world Benchmark} utilizes authentic, request-based data to test the model's adaptability and robustness under practical application conditions.

\vspace{-0.5em}
\begin{itemize}[topsep=5pt, leftmargin=1em]
    \item \textbf{Synthetic benchmark:} we propose a \textit{Synthetic Benchmark} based on the previous needle-based tasks NeedleInAHaystack~\cite{LLMTest_NeedleInAHaystack, li2024needlebenchllmsretrievalreasoning}, a standard methodology for information retrieval. This task performs fine-grained evaluations on the retrieval of specific sentences—referred to as "needles"—that are embedded across two dimensions: varying context lengths and differing depths within the same context. In addition to the original sets of needles from the NeedleInAHaystack benchmark, we designed three additional types of needles, targeting distinct task categories: multi-hop reasoning, multi-question, and multi-answer retrieval tasks.
    
    (\underline{i}) The multi-hop reasoning task draws inspiration from~\cite{ho2020constructing}. We developed distinct sets of needles that are interconnected through reasoning dependencies. 
    (\underline{ii}) The multi-question task, based on~\cite{son2024multi}, evaluates the model's capability to retrieve relevant information for multiple questions simultaneously. For this purpose, we created multiple questions, where the answer to each question corresponds to a specific subset of needles. Then, multiple (question, needles) pairs were randomly chosen to construct a multi-question task. 
    (\underline{iii}) The multi-answer task addresses limitations identified in previous studies~\cite{tang2023large}, which highlight the tendency of LLMs to exhibit "lazy retrieval". LLMs tend to only retrieve a subset of possible answers while neglecting others. To assess this, we designed tasks that require the model to extract all the correct answers for a single query. 
    Although our synthetic needle-based benchmark benefits from adhering to a widely recognized standard methodology, it has certain limitations. Specifically, the needles are inserted in irrelevant contexts, which isn't the case in a real usage scenario. 
    Another problem lies in the limited amount of manually designed sets of needles.

    \item \textbf{Open-source benchmark:} to address the limitations of \textit{Synthetic Benchmark}, we propose an \textit{Open-source Benchmark} derived from three open-source task-dedicated datasets, each corresponding to a specific task category. 
    Unlike retrieving the needles which are inserted in irrelevant background context, 
    this approach provides a more general assessment of retrieving information within context-relevant passages based on the questions. It also contains extensive test cases, enabling a comprehensive evaluation. To integrate these open-source datasets into our workflow, the original test cases of the datasets are reformatted into a tag-based structure. We introduce the construction of three categories of tasks as follows. (\underline{i}) Reasoning tasks are derived from Multi-hop QA Dataset~\cite{ho-etal-2020-constructing}. This dataset is designed to evaluate reasoning ability by challenging a model to analyze and combine information from multiple paragraphs to a given question. As introduced in the pre-processing Section~\ref{sec:preprocessing}, we enclose each sentence of context in this dataset within the \texttt{<TAG-${i}$>} and \texttt{</TAG-${i}$>} tag-pairs. The tags associated with the necessary reasoning sentences provided by the original Multi-hop QA Dataset are utilized as ground-truth tags for our Reasoning tasks. (\underline{ii}) Multi-answer tasks are made from MulTiple~\cite{2023} dataset. This dataset assesses the ability of LLMs to answer multiple questions simultaneously, ensuring no question is overlooked. The dataset includes two types of questions: question-dependent questions and document-dependent questions. The former explicitly specifies the number of demanded answers, while the latter challenges LLMs to determine the number of answers itself. For general purposes, document-dependent questions are selected as evaluation queries, assessing whether LLMs are able to retrieve all the answers associated with a given question. Document-dependent questions with auxiliary information of answer numbers are provided to GPT-4o to generate the ground truth. (\underline{iii}) Multi-question tasks are constructed from MTI BENCH~\cite{son2024multitaskinferencelargelanguage}. This benchmark is designed to evaluate LLMs' ability to handle multiple questions with one single inference. We first extract all the original test cases with (question, contents) pairs from the MTI BEBCH dataset, for which can assess the information retrieval ability. Then, we crafted 300 multi-question test cases by combining multiple extracted (question, contents) pairs together. The original dataset only provides the index of paragraphs where the answers are in, even though only a few sentences within the paragraph are truly demanded. Thus, we utilize GPT-4o to refine these answers, extracting only the target sentences while ignoring irrelevant parts, thereby providing more accurate and fine-grained ground truth tags for our multi-question tasks.
        
    \item \textbf{Real-world benchmark:}  we directly extract request-based tests from real-world data, focusing on assessing the models' general performance in application scenarios. We sampled $500$ pairs of data from real-world data sources, each consisting of a request and corresponding contents. Subsequently, those contents are tagged by the finite state machine described in Section~\ref{sec:preprocessing}. This approach tests our models under diverse realistic scenarios. By leveraging real-world data, the benchmark provides a more accurate measure of the model's robustness and adaptability compared with synthetic data.

\end{itemize}

\vspace{-0.15em}
\textbf{Evaluation Metrics.} To better align with our three benchmarks, we employ the following metrics regarding the relevance of the data. The data will be considered as \textit{positive relevant} if the context contains information relevant to the user query. Otherwise, it will be considered \textit{negative relevant}. All the test data in the \textit{Synthetic Benchmark} and \textit{Open-source Benchmark} is \textit{positive relevant}, since the tags for this evaluation corresponding to each needle are predetermined. More specifically, we extract ${<}\texttt{Summary}_{k_i},\texttt{TAG}_{k_i}>_{i=1,...,n} $ for each piece of evaluation data outlined in Section \ref{sec:instruct}. The test data adopted for request-based tests in the Real-world benchmark may be \textit{negative relevant}, as there is the possibility of the absence of ground truth.
we use \texttt{Precision}, \texttt{Recall}, and \texttt{F1} score, typical assessment methods for multi-class classification, as our key evaluation metrics.
We formulate the measurement of \textit{negative relevant} data as a multi-label matching problem, where \texttt{Exact Match Ratio} considers only fully correct matches of each \texttt{TAG}s, is particularly suitable for scenarios where the ground truth is None. By utilizing this property, the \texttt{Exact Match Ratio} is adopted to examine a model's capacity to reject irrelevant data, made-up, or inconsistent data.

\vspace{0.15em}
\textbf{Baseline Models.}
We select a set of baseline models for comparison, including open-source models such as \textsc{Qwen2-72B} and \textsc{Yi-9B-Chat}, along with the closed-source model \textsc{GPT-4o}. Additionally, we compare these models against our proposed \textsc{Extractor-LLM}.

\vspace{0.15em}
\textbf{Experiment Results and Discussion.} This subsection presents the experiment results on synthetic, open-source benchmarks, and real-world benchmarks, respectively. For the needle-based \textit{Synthetic Benchmark}, we calculate the average \texttt{F1} score across all results, considering every context length and insertion point depth for each needle.

\begin{itemize}[topsep=5pt, leftmargin=1em] 
    \item \textbf{Sythetic benchmark result:}
    The results of \textit{Sythetic Benchmark} tests are summarized in Table~\ref{tbl:test-table-1}. As one of the most powerful models, \textsc{GPT-4o} achieves leading performance across all task scenarios, ranking first in base, reasoning, and multi-answer retrieval tasks, and tying with \textsc{Extractor-LLM} in the multi-question tasks. While \textsc{Extractor-LLM} delivers slightly lower scores than \textsc{GPT-4o} in other three tasks, it remains a strong competitor, demonstrating robust and consistent performance across all tasks. These results highlight \textsc{Extractor-LLM}’s reliability in our context retrieval workflow, standing out as a versatile alternative just behind \textsc{GPT-4o} in key metrics.

        \begin{table}[h]
            \centering
            \caption{Average \texttt{F1} scores (higher is better) across different categories of tasks in \textit{Sythetic Benchmark}. "Reasoning" refers to "Reasoning tasks," "Multi-Ans" to "Multi-answer tasks," and "Multi-Q" to "Multi-question Tasks.}
            \begin{tabular}{lcccc}
            \toprule 
            Model & Base & Reasoning & Multi-Ans & Multi-Q \\
            \midrule
            \textsc{Yi-9B-Chat} & 0.328 & 0.147 & 0.284 & 0.166 \\
            \textsc{Qwen2-72B} & 0.732 & 0.346 & 0.675 & 0.946 \\
            \textsc{GPT-4o} & 0.969 & 0.904 & 0.986 & 1.0 \\
            \textsc{Extractor-LLM} & 0.925 & 0.803 & 0.973 & 1.0 \\
            \bottomrule
            \end{tabular}
            \label{tbl:test-table-1}
            \vspace{-0.75em}
        \end{table}

    \item \textbf{Open-source benchmark result:} The results of the \textit{Open-source Benchmark} tests are summarized in Table~\ref{tbl:benchmark-based-table-1}. \textsc{GPT-4o} demonstrates exceptional performance, achieving the highest score in the multi-question tasks. However, \textsc{Extractor-LLM} achieves a very close score in the multi-question tasks while surpassing all other models, including \textsc{GPT-4o}, in both reasoning and multi-answer tasks, showcasing its cutting-edge capability to handle complex reasoning and multi-answer extraction challenges with remarkable accuracy. Notably, other models such as \textsc{Yi-9B-Chat} and \textsc{Qwen2-72B} trail behind, further emphasizing the advancements achieved by \textsc{Extractor-LLM} and \textsc{GPT-4o}.

    \begin{table}[h]
            \centering
            \caption{Average \texttt{F1} scores (higher is better) across different categories of tasks in \textit{Open-source Benchmark}. "Multi-Ans"  refers to "Multi-answer tasks," and "Multi-Q" to "Multi-question Tasks."}
            \begin{tabular}{lcccc}
            \toprule 
            Model & Reasoning & Multi-Ans & Multi-Q \\
            \midrule
            \textsc{Yi-9B-Chat} & 0.361 & 0.328 & 0.439 \\
            \textsc{Qwen2-72B} & 0.568 & 0.441 & 0.570 \\
            \textsc{GPT-4o} & 0.591 & 0.536 & 0.755 \\
            \textsc{Extractor-LLM} & 0.724 & 0.581 & 0.687 \\
            \bottomrule
            \end{tabular}
            \label{tbl:benchmark-based-table-1}
        \end{table}

    \item \textbf{Real-world benchmark result:} 
    The results of \textit{Real-world benchmark} are summarized in Table~\ref{tbl:test-table-2}. \textsc{GPT-4o} delivers high performance across all metrics, slightly leading in \texttt{Recall} and achieving competitive scores in other areas. However, \textsc{Extractor-LLM} excels by surpassing \textsc{GPT-4o} in \texttt{Precision}, \texttt{F1} score, and \texttt{Exact Match (EM)}, highlighting its superior accuracy and robustness in real-world scenarios. In contrast, \textsc{Yi-9B-Chat}’s accuracy is particularly low because it tends to output a large portion of the \texttt{TAG}s from content indiscriminately, resulting in a high \texttt{Recall} but a very low \texttt{Precision} as well as a pretty low Exact Match Ratio. These results emphasize the consistent advancements of \textsc{Extractor-LLM} showcasing the strong context retrieval abilities under the real-world scenario.

        \begin{table}[h]
            \centering
            \caption{\textit{Real-world Benchmark} results across different models, evaluated using metrics including \texttt{Recall}, \texttt{Precision}, \texttt{F1}, and \texttt{Exact Match Ratio (EM)}, where higher values indicate better performance.}
            \begin{tabular}{lccccc}
            \toprule 
            Model & Recall & Precision & F1 & EM \\
            \midrule
            \textsc{Yi-9B-Chat} & 0.8383 & 0.4722 & 0.6041 & 0.1729 \\
            \textsc{Qwen2-72B} & 0.7145 & 0.6482 & 0.6797 & 0.7078 \\
            \textsc{GPT-4o} & 0.8357 & 0.6691 & 0.7431 & 0.7563 \\
            \textsc{Extractor-LLM} & 0.8353 & 0.7139 & 0.7698 & 0.7985 \\
            \bottomrule
            \end{tabular}
            \label{tbl:test-table-2}
            \vspace{-1.5em}
        \end{table}

\end{itemize}



\section{Related Work}
\label{sec:rel}

RAG is a powerful approach that significantly enhances the quality of LLM-generated content by retrieving and integrating relevant information from data sources not visited during the generative LLM training phase. In this section, we focus on the relevant techniques for Internet-level search augmented generation and model tuning techniques under the RAG paradigm instead of comprehensive RAG literature review~\cite{zhao2024retrieval,ding2024survey}.

\subsection{Search Augmented Generation}

Classic RAG systems leverage vector databases~\cite{cai2023bonsaikv,zhang2023experimental,lu2024fluidkv,yu2022treeline,ahmad2022pantheon,wang2023mirrorkv,mo2023learning,su2024vexless} to retrieve semantically similar chunks of text from a static corpus, which are usually limited by the scope of their pre-indexed datasets. On the other hand, 
Internet-level dynamic search-augmented generation~\cite{nakano2021webgpt,komeili2022internet,lazaridou2022internet,li2023web} leverages the most up-to-date information available online to significantly enhance the quality of question-answering~\cite{nakano2021webgpt,lazaridou2022internet,li2023web} and dialogue generation~\cite{komeili2022internet} within some specific domains. Furthermore, this approach effectively addresses the challenge of hallucinations in LLM-generated outputs~\cite{komeili2022internet,xie2024weknow} by grounding the generation process in up-to-date, verifiable data retrieved from the web. By integrating real-time information, these systems~\cite{nakano2021webgpt,komeili2022internet,lazaridou2022internet,li2023web} ensure that the responses are not only relevant and accurate but also reflective of the latest developments, thereby improving the overall reliability of the generative inference.   

\subsection{Model Tuning in RAG}

To enhance the quality of outputs generated by LLMs, various learning approaches and learnable components have been explored within the RAG paradigm to integrate retrieved information effectively~\cite{siriwardhana2023improving,luo2023search,yoranmaking}. For example, WebGPT~\cite{nakano2021webgpt} fine-tuned the generative LLM, i.e., GPT-3, to answer long-form questions by using a text-based web-browsing environment, enhancing both retrieval and synthesis in an end-to-end manner; recently, RankRAG~\cite{yu2024rankrag} instruction-tunes an end-to-end generative LLM for the dual purpose of context ranking and answer generation in RAG. 
Instead of re-training the generative LLM, 
DRAGIN~\cite{su2024dragin} introduces a module dynamically determining when and what to retrieve based on the LLM’s information needs during the text generation process; similarly, an advanced continual knowledge learning approach has been explored to selectively decide whether to access the Internet or not dynamically~\cite{li2023web}. 
Query re-write module~\cite{ma2023query,chan2024rq} has also been studied to eliminate ambiguity in the original input prompt; for example, RQ-RAG~\cite{chan2024rq} tuned a 7B plugin model in RAG paradigm to dynamically refine search queries through explicit rewriting, decomposition, and disambiguation. 


\section{Conclusion}
\label{sec:con}

In this paper, we have presented a novel paradigm, Internet search augmented generation for LLMs that can integrate real-time information through the use of standard search engine APIs, rather than relying on a static pre-processed corpus. This method addresses the limitations of the traditional RAG paradigm, which often fails to incorporate the most up-to-date information. 
Concretely, we implement a \textsc{parser-LLM} that determines whether Internet-augmented generation is necessary and, if so, extracts the relevant search keywords in a single inference pass; a \textit{mixed ranking strategy} that mitigates biases introduced by search engine APIs by re-ranking the retrieved HTML files; and an \textsc{extractor-LLM} designed to accurately and efficiently extract relevant information from the fresh content within each HTML file.
Our extensive empirical evaluations demonstrate that this Internet search augmented generation paradigm significantly enhances the quality of the generated content.
The successful deployment of this system in a production environment, specifically within 01.AI's generative inference framework, further validates the practical applicability of our approach.


\bibliographystyle{unsrt}
\bibliography{ref}

\clearpage




\end{document}